\documentstyle[12pt]{article}
\begin{document}
\thispagestyle{empty}
\setcounter{page}0

{\flushright
JINR preprint E2-94-254\\
Dubna, July 5, 1994\\}

\vfill
\begin{center}
{\Large\bf New Tensor Interactions in $\mu$ Decay} \vfill

{\large M. V. Chizhov}\footnote{E-mail: $physfac2@bgearn.bitnet$}
 \vspace{1cm}

{\em Bogoliubov Laboratory of Theoretical Physics, Joint Institute for
Nuclear Research, Dubna, Russia}\\ and \\ {\em
Center of Space Research and Technologies, Faculty of Physics,
University of Sofia, 1126 Sofia, Bulgaria}\footnote{Permanent address}

\end{center}  \vfill

\begin{abstract}

The most general form of the hamiltonian for the muon decay is
presented. We assume that it arises as a result of the exchange of
intermediate bosons with a momentum $q$ and naturally should depend on
this momentum. That allows us to introduce two additional coupling
constants for the tensor interactions which give rise to new parameters
in the energy spectrum of positrons. The experimental consequences of
such a generalization are discussed.

\end{abstract}

\vfill

{\em Submitted to ``Mod. Phys. Lett. A''}

\vfill

\newpage

\section{Introduction} \indent

One of the basic decay processes in the weak interactions is the $\mu$
decay $\mu^+\to e^+~\nu_e~\widetilde{\nu}_\mu$. Free of any QCD
complications, it can be used for a thorough check of the standard
theory of electroweak interactions and the determination of the Fermi
coupling constant $G_F$. The experimental accuracy is so high that the
one-loop electromagnetic radiative corrections must necessarily be taken
into account. Therefore, if some nonstandard interactions are of the
same order of magnitude, $O(\alpha)$, then they can be extracted on the
background of the V--A interaction and radiative corrections to it. In
the literature, the following most general form of the $\mu$-decay
hamiltonian is accepted \cite{Fetch}

\begin{equation}
{\cal H}={4 G_F \over \sqrt 2}
\sum_{\mbox{\tiny
 $\begin{array}{c} k=S,V,T\\ \epsilon,\chi=R,L\end{array}$}}
\Bigl\{ g^k_{\epsilon\chi}
 \left[ \bar{e}_{\epsilon} \Gamma^k \nu^e_n \right]
 \left[ \bar{\nu}^{\mu}_m \Gamma^k \mu_{\chi} \right] + {\rm h.c.}
\Bigr\} . \label{Ham}
\end{equation}

\noindent Here, $k$ labels the type of interaction (scalar, vector,
tensor), $\epsilon$ and $\chi$ indicate the chirality of the charged
leptons. (The chiralities of the neutrinos, $n$ and $m$, are uniquely
fixed by $\epsilon$, $\chi$, and $k$). The standard V--A interaction
implies that $g^V_{LL}$=1, and other $g^k_{\epsilon \chi}$ are zero.
Nonstandard couplings may arise in extensions of the standard model from
the exchange of new intermediate bosons, other than $W^{\pm}_{\mu}$. At
the first sight, it may seem that there are 12 (generally complex)
constants $g^k_{\epsilon\chi}$. However, the local tensor interactions
$[\bar{e}_R \sigma^{\alpha\beta}\nu^e_L][\bar{\nu}^{\mu}_{L}
\sigma_{\alpha\beta} \mu_R]$ and $[\bar{e}_L
\sigma^{\alpha\beta}\nu^e_R][\bar{\nu}^{\mu}_{R} \sigma_{\alpha\beta}
\mu_L]$ are identically equal to zero. Therefore, the coupling constants
$g^T_{RR}$ and $g^T_{LL}$ are absent in eq.(\ref{Ham}), and the most
general form of a local, Lorentz-invariant, derivative-free, and
lepton-number-conserving four-fermion interaction is parameterized by 10
model-independent constants.

In the present paper we are going to demonstrate that, abandoning the
locality of the {\em effective} Fermi interaction, we can introduce two
additional constants $g^T_{RR}$ and $g^T_{LL}$ in front of the
interaction terms which depend on the momentum transfer $q_{\mu}$. Such
terms arise from the exchange of tensor particles (the fundamental
interactions of which are {\em local}) in an extended model of the
electroweak interactions \cite{Chiz}. Let us redefine the tensor
structure to be

\begin{equation} \Gamma^T \otimes \Gamma^T \equiv {1 \over 2}
 \sigma^{\alpha \lambda} \otimes \sigma_{\beta\lambda} \cdot
 {4 q_\alpha q^\beta \over q^2} . \label{tensor}
\end{equation}

\noindent Then the terms with $g^T_{LR}$ and $g^T_{LR}$ in
eq.(\ref{Ham}) remain the same, owing to the identity

$$ \sigma^{\alpha \lambda} P_{\pm} \otimes \sigma_{\beta\lambda} P_{\pm}
\cdot {4 q_\alpha q^\beta \over q^2} = \sigma^{\alpha \beta} P_{\pm}
\otimes \sigma_{\alpha\beta} P_{\pm} , $$

\noindent where $P_{\pm}={1\over 2}(1\pm\gamma^5)$ is the chiral
projection operator. In the most general case, one can assume that the
effective four-fermion interaction arises from exchange of some bosons
with a momentum $q_\mu$ (that is, the interaction depends only on the
momentum transfer). Then, in fact, eq.(\ref{Ham}) with the above
definition (\ref{tensor}) will be the most general form of the effective
interaction of charged leptons (up to a factor depending on $q^2$) ---
the matrix structure $\gamma^\alpha \otimes \gamma^\beta \cdot q_\alpha
q_\beta$ for the particles on the mass shell is reduced to the scalar
structure $\Gamma^S \otimes \Gamma^S$.

The tensor particles have been introduced for the following reason. In
the recent experiments $\pi^- \to e^- ~ \widetilde{\nu} ~ \gamma $
\cite{Pi} and $K^+ \to \pi^{\rm o} ~ e^+ ~ \nu$ \cite{K}, tensor form
factors have been discovered. These form factors cannot be explained in
the framework of the standard V--A interaction \cite{Bel}. For the
semileptonic weak decays, an additional interaction has been introduced
\cite{Chiz}

\begin{equation}
{\cal{L}}_{qe}=-\sqrt 2~ G_F~ f_t~ \bar{u} \sigma^{\alpha\lambda} d'~
{q_{\alpha} q^{\beta} \over q^2}~ \bar{e}_R \sigma_{\beta\lambda} \nu_L
, \label{qe}
\end{equation}

\noindent where $d'= d \cos \theta_C + s \sin \theta_C$, and the value
of $f_t$ can be found from analyzing these meson decays. In the
framework of QCD, by applying the PCAC technique \cite{Bel}, we obtain
the value $f_t=(7.84\pm 2.24)\times 10^{-2}$ \cite{K0} from the
pion-decay data \cite{Pi}. If we assume that the coupling constants of
the tensor particles to quarks and leptons are the same, then we can
determine the tensor constants in eq.(\ref{Ham}):
$g^T_{LR}=g^T_{RL}=g^T_{LL}=0$~\footnote{The pion decay puts very strong
restrictions on the tensor couplings \cite{Vol}. Elimination of the
above three constants is enough to satisfy the constraints. That just
means that, as well as in the standard weak interactions, only
left-handed neutrinos take part in the tensor interactions.}

\begin{equation}
 g^T_{RR} = {f_t\over 4} = (1.96 \pm 0.56) \times 10^{-2} . \label{g}
\end{equation}

\noindent However, for the sake of generality, we consider the
model-independent case with all $g^T_{\epsilon\chi}\ne 0$. Introducing
the tensor coupling constants $g^T_{LL}$ and $g^T_{RR}$ leads to the
appearance of new parameters in the $e^+$ energy spectrum. Below we
discuss in detail the consequences for processing the experimental data
on decays of nonpolarized muons and for determining the Fermi constant.

\section{The positron energy spectrum} \indent

Generally speaking, one can assume the existence of right-handed
neutrinos. Since the upper bound on the muon neutrino mass is only
$m_{\nu_\mu}$$<$270~keV \cite{Data}, the effects of the neutrino mass
could be comparable to that of $m_e$. The mass of $\nu_e$ is always
neglected in this paper.

It is straightforward to calculate the $e^+$ energy spectrum for the
$\mu^+$ decay including all mass effects. Let us introduce the scaled
$e^+$ energy $x_e = 2E/(\omega m_\mu)$ which varies in the interval
$x_{\rm o}\le x_e \le 1$, where $\omega=1 +\epsilon_e^2
-\epsilon_\nu^2$, $x_{\rm o}= 2 \epsilon_e/\omega$, and
$\epsilon_e=m_e/m_\mu$, $\epsilon_\nu=m_\nu/m_\mu$. Then the spectrum
reads

\begin{eqnarray} {{\rm d}\Gamma \over {\rm d} x_e} &=&
{A~ G^2_F~ m^5_{\mu} \over 256 \pi^3}
\Bigl[ h_1 + {2 \over 9}\rho h_2 +
 4 \epsilon^2_e (1-\epsilon^2_{\nu}) \tau h_3 +
 \epsilon_e (\eta h_4 + \varepsilon h_5)
\Bigr. \nonumber \\*
&&
\Bigl. +~\epsilon_{\nu} (\lambda h_6 + 4 \epsilon^2_e \nu h_7) +
 \epsilon_e \epsilon_{\nu} (\sigma h_8 - \kappa h_5)
\Bigr] . \label{part}
\end{eqnarray}

\noindent The functions $h_1,...,h_8$ are given by

\begin{eqnarray*}
h_1 &=& \sqrt{x^2_e - x^2_{\rm o}}~ (1-x_e)^2~ {\omega^5 x_e \over u},
 \\
h_2 &=& \sqrt{x^2_e - x^2_{\rm o}}~ (1-x_e)^2~
\bigl\{ 4 u^3 - u^2 (5+5\epsilon^2_e+\epsilon^2_{\nu}) +
 u \left[ (1-\epsilon^2_e)^2-\epsilon^2_{\nu}(1+\epsilon^2_e) \right]
\bigr. \\*
&&
\bigl. +~2 \epsilon^2_{\nu}(1-\epsilon^2_e)^2
\bigr\} {\omega^4 \over u^3} , \\
h_3 &=& \omega^2 (2 x_e-1) \ln {t_{\max}\over t_{\min}} -
 \left[ 1 + {u \over u-p} \right]
 {\omega^3 (1-x_e) \sqrt{x^2_e-x^2_{\rm o}} \over u}, \\
h_4 &=& 2 \sqrt{x^2_e - x^2_{\rm o}}~ (1-x_e)^2~ {\omega^4 \over u}, \\
h_5 &=&
\left[ 1+\epsilon^2_e+\epsilon^2_\nu - 3u +
 {\epsilon^2_\nu (1-\epsilon^2_e) \over u}
\right] {\omega^3 (1-x_e) \sqrt{x^2_e-x^2_{\rm o}} \over u} -
2\epsilon^2_e (1-\epsilon^2_\nu) \ln {t_{\max}\over t_{\min}}, \\
h_6 &=& \sqrt{x^2_e - x^2_{\rm o}}~ (1-x_e)^2~
 {\omega^4 (1-\epsilon^2_e-u) \over u^2}, \\
h_7 &=& \epsilon^2_e \ln {t_{\max}\over t_{\min}} -
{\omega^3 (1-x_e) \sqrt{x^2_e - x^2_{\rm o}} \over u}, \\
h_8 &=& \sqrt{x^2_e - x^2_{\rm o}}~ (1-x_e)^2~
 {\omega^4 (1-\epsilon^2_e+u) \over u^2},
\end{eqnarray*}

\noindent where $u=1 +\epsilon_e^2 -\omega x_e$, $p=\epsilon^2_\nu
(1-\epsilon^2_e-\epsilon^2_\nu) / (1-\epsilon^2_\nu)$, and

$$
t^{\max}_{\min} = {1 \over 2u}
\left[ u (1+\epsilon^2_e+\epsilon^2_\nu) -u^2-
 \epsilon^2_\nu (1-\epsilon^2_e) \pm (u-\epsilon^2_\nu)
 \sqrt{(1-\epsilon^2_e)^2-2u(1+\epsilon^2_e)+u^2}
\right] . $$

The Michel parameter $\rho$ and the quantities $\eta$, $\lambda$,
$\sigma$, $\tau$, $\varepsilon$, $\nu$, and $\kappa$ are functions of
the coupling constants $g^V_{\epsilon\chi}$:

\begin{eqnarray*}
\rho &=& {3\over A}
\left\{ \sum_{\epsilon,\chi=R,L}
 \left| g^S_{\epsilon\chi} -2 g^T_{\epsilon\chi} \right| ^2
 +4 \left| g^V_{LL} \right| ^2 +4 \left| g^V_{RR} \right| ^2
\right\} , \\
\eta &=& {8\over A} \sum_{\epsilon,\chi=R,L} {\rm Re}
\left\{ g^V_{\epsilon\chi}
 \left( g^S_{\bar{\epsilon}\bar{\chi}} +6 g^T_{\bar{\epsilon}\bar{\chi}}
 \right) ^*
\right\} , \\
\lambda &=& {8 \over A}~ {\rm Re}
\left\{ g^S_{LL} g^{S*}_{LR}
 +g^S_{RR} g^{S*}_{RL}
 -2 g^V_{LL} g^{V*}_{LR}
 -2 g^V_{RR} g^{V*}_{RL}
 -4 g^T_{LL} g^{T*}_{LR}
 -4 g^T_{RR} g^{T*}_{RL}
\right\} , \\
\sigma &=& {8\over A} \sum_{\epsilon,\chi=R,L} {\rm Re}
\left\{ g^V_{\epsilon\chi}
 \left( g^S_{\bar{\epsilon}\chi} -6 g^T_{\bar{\epsilon}\chi}
 \right) ^*
\right\} , \\
\tau &=& {16\over A} ~
\left\{
 \left| g^T_{LL} \right| ^2 + \left| g^T_{RR} \right| ^2
\right\} , \\
\varepsilon &=& {32 \over A}~ {\rm Re}
\left\{ g^V_{LL} g^{T*}_{RR}
 +g^V_{RR} g^{T*}_{LL}
\right\} , \\
\nu &=& {32 \over A}~ {\rm Re}
\left\{ g^T_{LL} g^{T*}_{LR}
 +g^T_{RR} g^{T*}_{RL}
\right\} , \\
\kappa &=& {32 \over A}~ {\rm Re}
\left\{ g^V_{LR} g^{T*}_{RR}
 +g^V_{RL} g^{T*}_{LL}
\right\} ,
\end{eqnarray*}

\noindent where the bar denotes opposite chiralities, and
$$ A = 4 \sum_{\epsilon,\chi=R,L}
 \left\{ \left| g^S_{\epsilon\chi} \right| ^2
  +4 \left| g^V_{\epsilon\chi} \right| ^2
  +12 \left| g^T_{\epsilon\chi} \right| ^2
 \right\} . $$

Notice that introducing new tensor constants leads to the appearance of
new structures $\tau$, $\varepsilon$, $\nu$, and $\kappa$ in the
spectrum. This also gives rise to a change in the definition of the
quantities $\rho$, $\eta$, $\lambda$, and $\sigma$ in favor of a more
symmetric form as compared to their standard definition \cite{Gr}.

Integrating over the whole spectrum, we can derive the partial decay
width of the muon into a positron

\begin{equation} \Gamma =
 {A~ G_F^2~ m_\mu^5 \over 256 \pi^3}
 \left[ H_1 + \epsilon_e \eta H_4
  +\epsilon_\nu ( \lambda H_6 +4 \nu \epsilon_e^2 H_7 )
  +\epsilon_e \epsilon_\nu \sigma H_8
 \right] , \label{width}
\end{equation}

\noindent where $H_i=\int _{x_{\rm o}}^1 h_i(x_e) {\rm d} x_e$
($i$=1,...,8).
\begin{eqnarray*}
H_1 &=& {1 \over 12} R_{\rm o} (1 - 7\epsilon_e^2 - 7\epsilon_\nu^2
 - 7\epsilon_e^4 - 7\epsilon_\nu^4 + 12\epsilon_e^2 \epsilon_\nu^2
 + \epsilon_e^6 + \epsilon_\nu^6
 - 7\epsilon_e^4 \epsilon_\nu^2 - 7\epsilon_e^2 \epsilon_\nu^4) \\*
&& +2 \epsilon_e^4 (1 - \epsilon_\nu^4) L_e
 +2 \epsilon_\nu^4 (1 - \epsilon_e^4) L_\nu , \\
H_2 &=& H_3~=~0, \\
H_4 &=& {1\over 3} R_{\rm o} (1 +10\epsilon_e^2 -5\epsilon_\nu^2
 +\epsilon_e^4 -2\epsilon_\nu^4 -5\epsilon_e^2\epsilon_\nu^2 )
 -4\epsilon_e^2 \left[ \epsilon_e^2 +(1 -\epsilon_\nu^2)^2 \right] L_e
 \\*
&& +4\epsilon_\nu^4 (1 -\epsilon_e^2 ) L_\nu, \\
H_5 &=& 0, \\
H_6 &=& {1\over 3} R_{\rm o} (1 -5\epsilon_e^2 +10\epsilon_\nu^2
 -2\epsilon_e^4 +\epsilon_\nu^4 -5\epsilon_e^2\epsilon_\nu^2 )
 +4\epsilon_e^4 (1 -\epsilon_\nu^2 ) L_e \\*
&& -4\epsilon_\nu^2
 \left[ \epsilon_\nu^2 +(1 -\epsilon_e^2)^2 \right] L_\nu , \\
H_7 &=& - {1\over 2} R_{\rm o} (1 +5\epsilon_e^2 +\epsilon_\nu^2 )
 +2\epsilon_e^2
 \left[ 2 (1 -\epsilon_\nu^2 )
  +{ \epsilon_e^2 (1 +\epsilon_\nu^2) \over 1 -\epsilon_\nu^2 }
 \right] L_e \\*
&& +2\epsilon_\nu^2
 \left[ 1 -2\epsilon_e^2
  +{ \epsilon_e^4 \over 1 -\epsilon_\nu^2 }
 \right] L_\nu , \\
H_8 &=& {1\over 3} R_{\rm o} (2 +5\epsilon_e^2 +5\epsilon_\nu^2
 -\epsilon_e^4 -\epsilon_\nu^4 -10\epsilon_e^2\epsilon_\nu^2 )
 -4\epsilon_e^2
 \left[ \epsilon_e^2\epsilon_\nu^2 +(1-\epsilon_\nu^2)^2 \right] L_e \\*
&& -4\epsilon_\nu^2
 \left[ \epsilon_e^2\epsilon_\nu^2 +(1-\epsilon_e^2)^2 \right] L_\nu ,
\end{eqnarray*}
where
$$ L_e=\ln {1+\epsilon_e^2-\epsilon_\nu^2+R_{\rm o} \over
 2 \epsilon_e}, ~~~~~~~~~~
L_\nu=\ln {1-\epsilon_e^2+\epsilon_\nu^2+R_{\rm o} \over
 2 \epsilon_\nu},$$
and $R_{\rm o}=\sqrt{(1-\epsilon_e^2)^2
-2\epsilon_\nu^2 (1+\epsilon_e^2) +\epsilon_\nu^4}$.

As one should have expected, the partial decay width (\ref{width}) does
not depend on the Michel parameter $\rho$ as well as on $\tau$,
$\varepsilon$, and $\kappa$, because $H_2=H_3=H_5=0$.

\section{Conclusions} \indent

To most clearly represent the effect of the new tensor constant, we
neglect the masses of the neutrinos and the positron, and explicitly
extract the dependence of the positron energy spectrum on $g^T_{RR}$ by
setting $g^V_{LL}$=1. Then

\begin{eqnarray} {\rm d} \Gamma &\propto &
\left\{ (1-x_e) + {2 \over 9}\rho_{\rm o} (4 x_e-3) + 2\epsilon_e
 \left( {1-x_e \over x_e} \eta_{\rm o} + {1 \over x_e} g^T_{RR}
 \right)
 + {(g^T_{RR})^2 \over 6} (15-14x_e)
\right. \nonumber \\*
&&
\left. + {3\alpha \over \pi} f(x_e)
\right\} x^2_e~ {\rm d} x_e , \label{prop}
\end{eqnarray}

\noindent where $\rho_{\rm o}$, $\eta_{\rm o}$, and $g^T_{RR}$ can be
considered as model-independent parameters, and $f(x_e)$ is a known
function which describes the one-loop electromagnetic radiative
correction \cite{Kino}. Taking eq.(\ref{g}) into account, we see that
the contribution of the new terms is of the same order of magnitude as
the one-loop electromagnetic correction. A more precise measurement of
the energy spectrum would allow one to detect this contribution.

If we assume that all the constants but $g^V_{LL}$ and $g^T_{LL}$ are
equal to zero, and the presented above theoretical curve (\ref{prop})
adequately describes the experiment, then the fit of the parameter
$\rho$, neglecting the tensor contributions, should lead to a systematic
deviation $\delta\rho=\rho-0.75$:

\begin{equation} \delta\rho = {9 \over 64}~
 {\strut \displaystyle \sum_{i=0}^n x_i^4 (x_i - 0.75)
  \left[ \epsilon_e {g^T_{RR} \over x_i}
   +{(g^T_{RR})^2 \over 6} (15 -14 x_i)
  \right]
 \over \strut \displaystyle \sum_{i=0}^n x_i^4 (x_i - 0.75)^2} .
\end{equation}

\noindent It is evident that the fit over the low-energy positrons
$x_e<0.75$ leads to $\delta\rho_{low}<0$ while for the high-energy
part of the spectrum $x_e>0.75$ one obtains $\delta\rho_{high}>0$. A
diffident indication of the existence of an effect can be observed
already in the data of ref.\cite{Sher}. If the experimental errors were
less, we would assert more assuredly that $\rho_{low} < \rho_{high}$.

The partial decay width derived from eq.(\ref{prop}) at $\eta_{\rm o}=0$
\begin{equation} \Gamma = {G_F^2 m_\mu^5 \over 192 \pi^3}
 \left[ 1 +12\epsilon_e g^T_{RR} +3 (g^T_{RR})^2 \right]
 \left[ 1 - {\alpha \over 2 \pi} \left( \pi^2 - {25 \over 4} \right)
 \right]
\end{equation}

\noindent can be used for evaluating the Fermi constant $G_F$. Notice
that if the value of $g^T_{RR}$ is given by eq.(\ref{g}), then the
contribution of the new interaction is comparable to the one-loop
electromagnetic radiative correction. This may lead to a perceptible
change in the value of $G_F$. \vspace{1cm}

\pagebreak[3]
{\bf Acknowledgements} \vspace{4mm}

I am grateful to L.V.Avdeev, D.I.Kazakov, O.V.Seliugin and V.I.Selivanov
for helpful discussions. I acknowledge the hospitality of the Bogoliubov
Laboratory of Theoretical Physics, JINR, Dubna, where this work has been
completed. The work is financially supported by Grant-in-Aid for
Scientific Research F-214/2096 from the Bulgarian Ministry of Education,
Science and Culture.

\pagebreak[3]

\end{document}